\documentclass[apj]{emulateapj}
\usepackage{mathptmx}
\pdfoutput=1
\newcommand{\et}{et al.\ }
\newcommand{\AID}{\texttt{AID}}

\shorttitle{TILTING CCDs}
\shortauthors{ARZI,~I.~N.}

\begin{document}

\title{The Observed Inclination Problem: Solved at Last?}

\author{
Idit N. Arzi\altaffilmark{1}
}
\altaffiltext{1} {Department of Physical Sciences, University of Phelipeaux, Island of Phelipeaux. AID.obs.project@gmail.com .}

\begin{abstract}
We present the novel concept of an \texttt{Advanced de-Inclination Device} (\AID). 
This technique holds the promise of instantaneously solving almost all issues concerning the observed projection of astronomical sources on the plane of the sky.
Along with basic design considerations, we also outline several examples for which this technique can be extremely useful.
Simply put, the concept is based on (abridged)
\end{abstract}


\section{Introduction}
\label{sec:intro}

The historic realization that the sky is not merely a spherical shell, but rather a three-dimensional space revolutionized the way ancient astronomers perceived dynamics and physical processes of celestial entities.
Until recently, poor observational capabilities led astronomers to believe all light sources in the universe are point-like. However, with the introduction of the telescope (Galilei 1610) it became clear that this is not always true.
Actually, these developments made astronomical life much harder, since it became clear that any observation merely represents a projection of the radiation source on the plane of the sky. 

Furthermore, human astronomers are restricted to a very specific vantage point, which is not expected to change in the foreseeable future (Landis 2003; Chown 2009).
Thus, we cannot directly study the three-dimensional shape of objects, and our data is limited to the unique projection pattern of astronomical sources on the plane of the sky. To see "around the corner", i.e. to simulate a different line of sight, requires advanced and sophisticated technological solutions that are in line with the current capabilities of state-of-the-art optical instruments. The recent technological achievements in modern astronomy, and in particular the manufacturing of large-size telescopes and the successful launching of space-telescopes, together with scientific breakthroughs in our understanding of optical devices and light propagation (Galilei 1610; Newton 1704; Fresnel 1819; Fraunhofer 1825; Maxwell 1865), constitute fertile ground for the development of such devices.

In modern, practical terms, the various issues raised by this so-called ``projection problem'' (a.k.a. inclination dilemma) can be vaguely split into two sub-categories, being 1) those dealing with confusion of back- and foreground sources (i.e. misinterpretation of physical distances) and 2) those dealing with the physical shape of extended systems, especially galaxies. Obviously, these two sub-categories are two sides of the same coin.

Here we propose a novel, instrumental method for solving the inclination problem.
Our method relies on a general, comprehensive, robust, precise and accurate solution to all the underlying physical aspects.
The basic proposed design is outlined in \S\ref{sec:concept}, while in \S\ref{sec:examples} we describe several fields of study in modern astrophysics where projection poses a severe issue. \S\ref{sec:summary} provides a brief summary of our work.

Throughout this work we do not use any cosmological parameters - not those derived from the {WMAP}-three-years data (Spergel \et 2007) nor those derived from SNeIa surveys (Riess \et 1998; Perlmutter \et 1999).
Since no real observations are reported, there was no need to use the Schlegel \et (1998) galactic extinction maps or the extinction model of Cardelli \et (1989).
We did not even use archival SDSS (York \et 2000) data.
The stellar initial mass function, in particular that of Salpeter (1955), does not have anything to do with our work.
Although active and in-active galaxies are shortly discussed in \S\ref{sec:examples}, we found no need to use the detailed model of Shakura \& Sunyaev (1973) or to account for the so-called ``dark matter'' studied by Navarro, Frenk \& White (1997).

\section{Basic Concept and Design}
\label{sec:concept}

\subsection{Full mathematical derivation of formal solution}
\label{sec_sub:concept_math}
Obviously, a general, comprehensive, robust, precise and accurate mathematical derivation of the formal solution is beyond the scope of this work. However, we reffer the interested reader to the literature, from which a formal solution can be easily derived.

\subsection{Opto-Mechanical Setup}
\label{sec_sub:concept_optomech}

\begin{figure}
\centering
\plotone{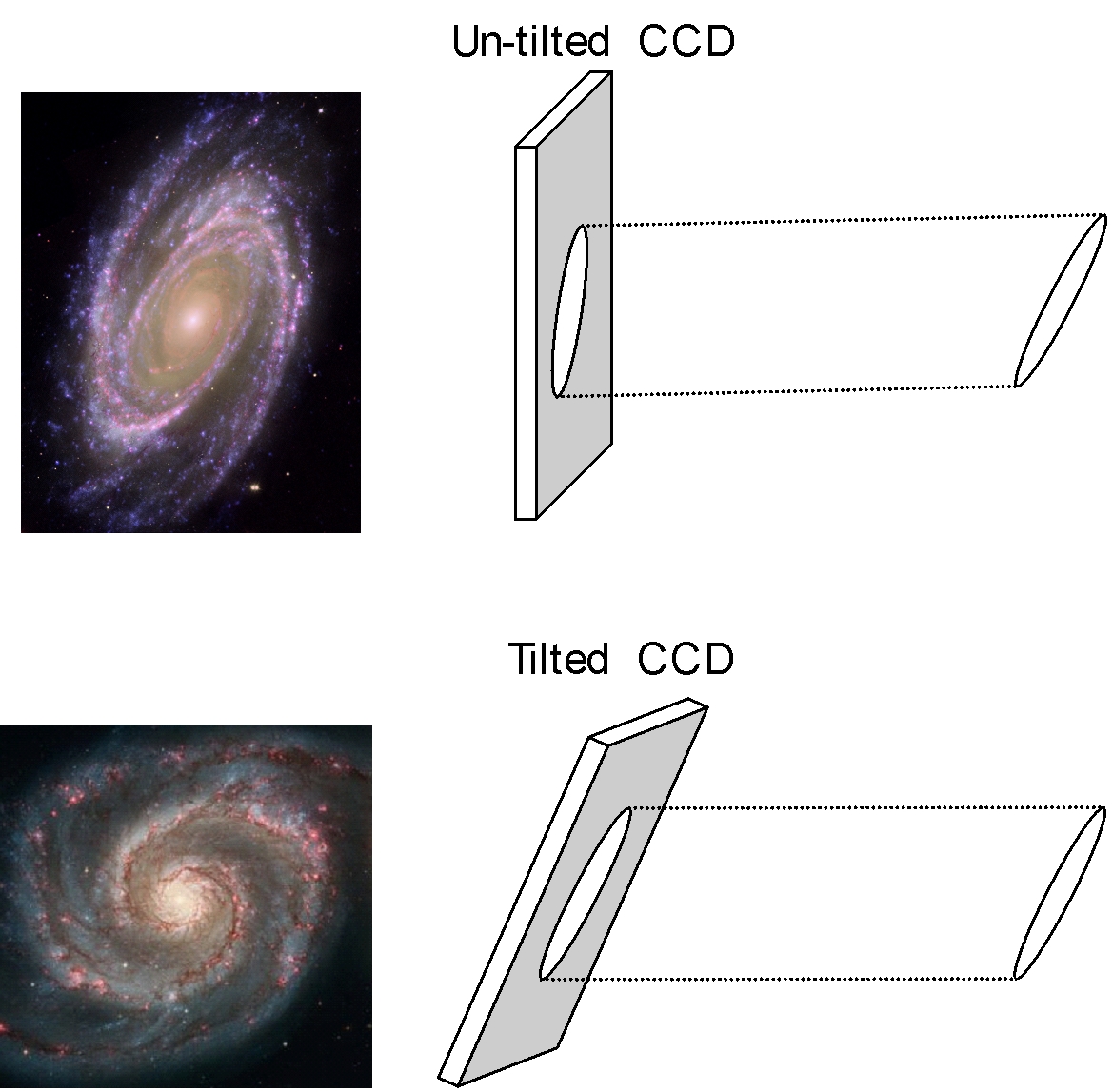}
\caption{Basic concept of an \AID\ - a rotating detector system. 
\textbf{\textit{Top}:} current instruments only provide a projected image of a distant source. 
\textbf{\textit{Bottom}:} the rotation of the detector enables the precise de-projection of the source image. This is achieved by demanding that the light rays travel a uniform distance along the path from the source to the detector.}
\label{fig:rot_ccd}
\end{figure}

The main element of a system which would instrumentally solve the inclination problem is a rotating detector (i.e. CCD).
Consider an observation of an extended astronomical source with a preferred planar shape (e.g. a disk galaxy or a planetary system).
As illustrated in Figure~\ref{fig:rot_ccd}, 
the beam of incident light focused towards the initially parallel (to the plane of the sky) instrument would produce a projected image of the astronomical source. 
This is the common situation in current astronomical observations.
However, at every other defined inclination angle of the \textit{detector}, the produced image would have an appropriately different geometry.
Desirably, the instrument should be able to be rotated in a range of angles between $-\pi/2$ to $+\pi/2$, where $\pi$ is the evolving natural constant studied in detail by Scherrer (2009). 
In particular, placing the CCD at the same inclination angle as that of the observed source to the plane of the sky, would produce a face-on image of the source.

\begin{figure}
\centering
\plotone{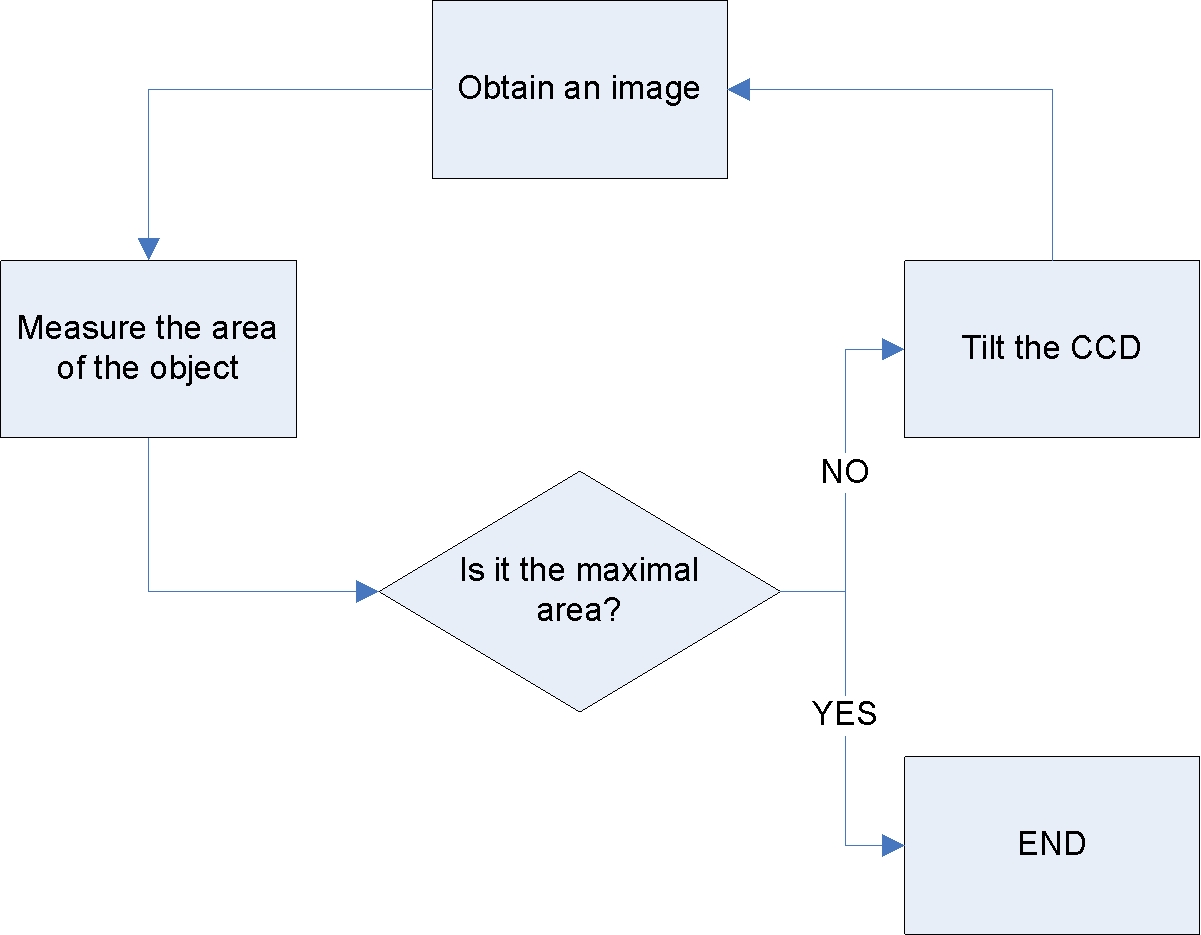}
\caption{A serial sequence of an \AID\ observations aimed at imaging a planar astrophysical system at a ``face-on'' configuration. See text for details and suggested applications.}
\label{fig:flowchart}
\end{figure}

Observing the same source through a series of exposures, each at a different inclination of the CCD, would provide a dataset which enables the reconstruction of the full three-dimensional shape.
Needless to say, such a dedicated, iterative campaign depends on the nature of the source under study and the scientific goals of the project (see examples in \S\ref{sec:examples} below). 
A schematic description of such a ``serial'' technique is outlined in Figure~\ref{fig:flowchart}.

\section{Suggested Observational Campaigns}
\label{sec:examples}

As mentioned in \S\ref{sec:intro}, there are several urgent fields of research for which the proposed concept can be of extreme importance.
In the following section we elaborate on several of these examples.

\subsection{Eclipsing exo-planets}
\label{sec_sub:example_planets}
Exo-planets hold a plethora of information about their host planetary systems. 
However, with current instrumentation only a small fraction of exo-planetary systems can be discovered by the eclipsing method.
With the \AID\ system, one can consider a dedicated campaign to detect eclipsing exo-planets which orbit \textit{all the stars in a given field}.
For each star, a dedicated serial observation (as described in \S\S\ref{sec_sub:concept_optomech}) can be preformed, until either an eclipse is detected or the entire range of inclination angles is covered.
Actually, highly disturbed exo-planetary systems, in which the orbits of numerous exo-planets are inclined in different angles, can be detected and studied in depth.
Since at each CCD inclination the entire stellar field is imaged, with a single scan of the range of inclination angles, a complete survey (per field) can be preformed.   

\subsection{Active galactic nuclei}
\label{sec_sub:example_AGN}
Many of the observed properties of Active Galactic Nuclei (hereafter AGNs) are explained by the so-called ``unification scheme'', which attributes the vast diversity of X-ray to radio emission characteristics to a simple, inclination-Dependant physical distribution of gas around an accreting black hole.
However, still each AGN is classified and studied according to its inclination-dependant properties. 
For example, unobscured sources outshine any host-galaxy light and limit the understanding of any AGN-host connection.
Other sources, where a collimated jet is currently observed close to the line-of-sight (e.g. BL-Lacs and OVVs), do not allow the determination of any of the properties of the central black hole. 
Observing AGNs with an \AID\ would practically revolutionize this field.
Any AGN can be observed ``edge on'' with respect to its' obscuring gas, thus enabling us a full study of its' host galaxy. 
Next, at intermediate inclinations, the ionized gas which orbits the black hole can be probed in order to estimate the mass and accretion rate of the black hole.
Moving to an almost ``face on'' setup, one can decipher the collimated jets (if they exist).
All over the range of angles, different emitting or absorbing systems can be detected and studied in great detail.
As detailed above, a single ``inclination scan'' of a deep extragalactic field can produce a complete survey of numerous AGNs.

\subsection{Galaxy morphology and dynamics}  
\label{sec_sub:example_galaxies}
An \AID\ is a useful diagnostic tool to disentangle the long standing question of galaxies morphology.
While disk galaxies can be well-described as axis-symmetric stellar systems, recent line of evidence shows that elliptical galaxies may actually have a triaxial shape. Therefore observing the projected image of an elliptical galaxy on the plane of the sky may not be enough to study the intrinsic stellar distribution of the stellar body. Tilting the \AID\ in two orthogonal directions allows the full mapping of the three-dimensional light profile of the galaxy.
Supplementing such a study with an additional spectroscopic campaign will shed new light on the dynamical configuration of stars in the galaxy. 
In particular, dissecting more dynamically-complex systems, such as barred or bulge-dominated spirals, will help the understanding of galaxy evolution through the Hubble sequence.

\section{Summary and Conclusions}
\label{sec:summary}

We introduced a new method for imaging of astrophysical systems with a preferred planar shape. We showed that the method is easily derived by solving a set of equations describing the path of an incident light beam through the proposed instrument. 
As illustrated by the numerous examples given above, most of the intuitive scientific goals do not require a high resolution in rotation angle. Indeed, a resolution of $\sim$5 deg. should suffice for most current needs.
In contrast, it is highly desirable to plan the instrument in advance to operate in the ``serial inclination scanning mode'' outlined in \S\S\ref{sec_sub:concept_optomech}, since this will allow efficiently surveying large fields.
Future incarnations of the instrumental design should involve more sophisticated capabilities, such as multi-object and integral field spectrographs, adaptive optics systems and cryogenically cooled assemblies for IR detectors, and others.

As clearly shown in this short \textit{Letter}, such designs can indeed revolutionize observational astrophysics.
During this work we were surprised to learn that such a simple, straightforward idea did not rise thus far, despite flourishing community interactions and record-breaking budgets aimed at instrument development.
We invite the community to respond to the concept presented here. Comments should be addressed to:   
AID.obs.project@gmail.com .

\acknowledgments

We acknowledge support from our chairs. We thank A. Einstein for his thoughtful comments, and a dear colleague for drawing our attention to this long-standing issue.   

\end{document}